\documentclass[journal=nalefd,manuscript=letter,layout=traditional]{achemso}

\usepackage{graphicx}
\usepackage{bm}
\usepackage{color}
\usepackage{amsfonts}
\usepackage{isomath}
\usepackage{amsmath}
\usepackage{amsthm}
\usepackage{amsfonts}
\usepackage{amssymb}
\usepackage{fdsymbol}
\usepackage{braket}
\usepackage{gensymb}

\setkeys{acs}{articletitle = true}

\title{Van der Waals engineering of ultrafast carrier dynamics in magnetic heterostructures}
\author{Paulina Ewa Majchrzak}
\affiliation{Department of Physics and Astronomy, Interdisciplinary Nanoscience Center, Aarhus University,
8000 Aarhus C, Denmark}
\author{Yuntian Liu}
\affiliation{Department of Physics and Shenzhen Institute for Quantum Science and Engineering
(SIQSE), Southern University of Science and Technology, Shenzhen 518055, China}
\author{Klara Volckaert}
\affiliation{Department of Physics and Astronomy, Interdisciplinary Nanoscience Center, Aarhus University,
8000 Aarhus C, Denmark}
\author{Deepnarayan Biswas}
\affiliation{Department of Physics and Astronomy, Interdisciplinary Nanoscience Center, Aarhus University,
8000 Aarhus C, Denmark}
\author{Chakradhar Sahoo}
\affiliation{Department of Physics and Astronomy, Interdisciplinary Nanoscience Center, Aarhus University,
8000 Aarhus C, Denmark}
\author{Denny Puntel}
\affiliation{Dipartimento di Fisica, Università degli Studi di Trieste, 34127 Trieste, Italy}
\author{Wibke Bronsch}
\affiliation{Elettra - Sincrotrone Trieste S.C.p.A., 34149 Basovizza, Italy}
\author{Manuel Tuniz}
\affiliation{Dipartimento di Fisica, Università degli Studi di Trieste, 34127 Trieste, Italy}
\author{Federico Cilento}
\affiliation{Elettra - Sincrotrone Trieste S.C.p.A., 34149 Basovizza, Italy}
\author{Xing-Chen Pan}
\affiliation{Advanced Institute for Materials Research, Tohoku University, Sendai 980-8577, Japan}
\author{Qihang Liu}
\affiliation{Department of Physics and Shenzhen Institute for Quantum Science and Engineering
(SIQSE), Southern University of Science and Technology, Shenzhen 518055, China}
\author{Yong P. Chen}
\affiliation{Department of Physics and Astronomy, Interdisciplinary Nanoscience Center, Aarhus University,
8000 Aarhus C, Denmark}
\alsoaffiliation{Advanced Institute for Materials Research, Tohoku University, Sendai 980-8577, Japan}
\alsoaffiliation{Department of Physics and Astronomy and School of Electrical and Computer Engineering and Purdue Quantum Science and Engineering Institute, Purdue University, West Lafayette, IN 47907, USA}
\author{S{\o}ren~Ulstrup}
\email{ulstrup@phys.au.dk}  
\affiliation{Department of Physics and Astronomy, Interdisciplinary Nanoscience Center, Aarhus University,
8000 Aarhus C, Denmark}

\begin{document}

\newpage

\begin{abstract}
Heterostructures composed of the intrinsic magnetic topological insulator MnBi$_2$Te$_4$ and its non-magnetic counterpart Bi$_2$Te$_3$ host distinct surface electronic band structures depending on the stacking order and exposed termination. Here, we probe the ultrafast dynamical response of MnBi$_2$Te$_4$ and MnBi$_4$Te$_7$ following near-infrared optical excitation using time- and angle-resolved photoemission spectroscopy, and disentangle surface from bulk dynamics based on density functional theory slab calculations of the surface-projected electronic structure.  We gain access to the out-of-equilibrium charge carrier populations of both MnBi$_2$Te$_4$ and Bi$_2$Te$_3$ surface terminations of MnBi$_4$Te$_7$,  revealing an instantaneous occupation of states associated with the Bi$_2$Te$_3$ surface layer followed by carrier extraction into the adjacent MnBi$_2$Te$_4$ layers with a laser fluence-tunable delay of up to 350 fs.  The ensuing thermal relaxation processes are driven by phonon scattering with significantly slower relaxation times in the magnetic MnBi$_2$Te$_4$ septuple layers. The observed competition between interlayer charge transfer and intralayer phonon scattering demonstrates a method to control ultrafast charge transfer processes in MnBi$_2$Te$_4$-based van der Waals compounds.\\
\\
KEYWORDS: Magnetic topological insulators,  MnBi$_2$Te$_4$,  van der Waals heterostructures, ultrafast carrier dynamics, time- and angle-resolved photoemission spectroscopy,  density functional theory.
\end{abstract}

\maketitle
Heterojunctions composed of two-dimensional (2D) semiconducting materials are highly promising for developing novel optoelectronic devices due to tunable band gaps and charge carrier dynamics \cite{Jin:2018}.  The interplay of charge and phonon interactions has been intensively examined in transition metal dichalcogenide heterostructures, unveiling sub 50-fs interlayer carrier extraction that is ideal for light harvesting and photovoltaics \cite{Hong:2014,Zheng:2017}. Van der Waals compounds based on the antiferromagnetic topological insulator MnBi$_2$Te$_4$ present another attractive candidate for advanced opto- and spintronics applications due to the realization of exotic phases such as the quantum anomalous Hall effect and axion electrodynamics \cite{gong2019experimental,otrokov2019unique,zhang2019topological,li2019intrinsic,otrokov2019prediction,Yujun2020,Liu2020,zhao2021routes,gu2021spectral}. Yet, optically-induced ultrafast carrier dynamics and charge transfer processes in these materials remain largely unexplored.  Moreover, the properties of these materials can be elegantly engineered via van der Waals stacking. For example,  combining a quintuple layer (QL) of the non-magnetic topological insulator Bi$_2$Te$_3$ with a septuple layer (SL) of MnBi$_2$Te$_4$  reduces the antiferromagnetic exchange interaction between neighbouring planes of Mn atoms, while strengthening the affinity of the system to undergo a spin-flop transition to ferromagnetic order \cite{Deng2021}. As a result, van der Waals heterostructures of the form MnBi$_2$Te$_4$(Bi$_2$Te$_3$)$_m$ ($m = 1, 2, 3, \ldots$) offer a platform for exploring the interplay between the magnetic properties and the topological phases of matter \cite{Otrokov2017,sun2019rational,wu2019natural,Xu2020,Hu2020,wu2020distinct,hu2020van,lu2021half}. 

In order to gain access to the energy- and momentum-resolved ultrafast carrier dynamics of MnBi$_2$Te$_4$(Bi$_2$Te$_3$)$_m$ heterostructures, we employ time- and angle-resolved photoemission spectroscopy (TR-ARPES). This approach utilizes a near-infrared optical excitation to induce direct optical transitions  from the occupied to the unoccupied states. The out-of-equilibrium charge carrier populations in the excited states are then measured by a time-delayed ultraviolet probe pulse,  giving direct insights into charge transfer processes and phonon relaxation between bulk and surface states on a given heterostructure termination \cite{Yan:2021_sciinst}. This approach has revealed thermal dynamics with a duration of 4~ps in MnBi$_2$Te$_4$ across the magnetic ordering temperature \cite{Nevola2020},  as well as photoinduced filling of the surface state hybridization gap of the QL termination in a heterostructure with stoichiometry $m=3$ \cite{Zhong2021}.  Here,  we determine how carrier extraction and relaxation is controlled by the balance of inter- and intralayer interactions by comparing the time-dependent evolution of excited electronic states between the $m=0$ and $m=1$ heterostructures,  i.e.  MnBi$_2$Te$_4$ and MnBi$_4$Te$_7$.  We observe an excited state population in the QL of MnBi$_4$Te$_7$ that is generated via our infrared pump pulse. This population is then gradually extracted into the SL with a fluence-dependent delay of up to 350~fs.  Thermal relaxation then proceeds at a slower rate within the magnetic SL layer, implying that the layer-dependent electronic and magnetic properties can be optically controlled on an ultrafast timescale.

\begin{figure} [t!]
	\includegraphics[width=0.98\textwidth]{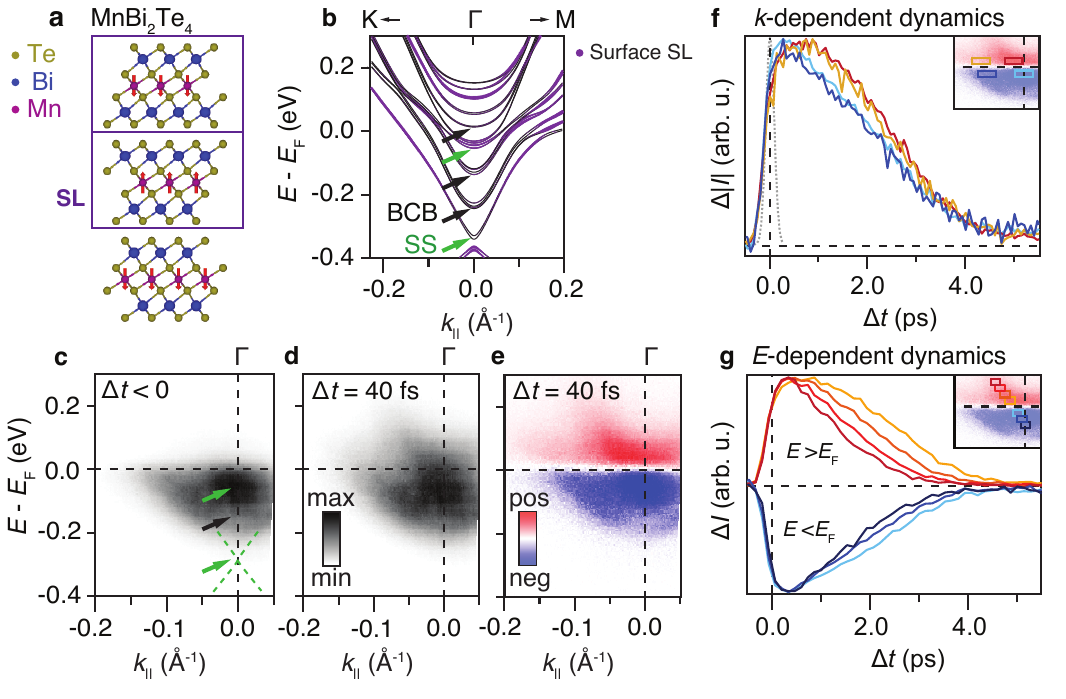}\\
\caption{Surface electronic structure and excited state populations in MnBi$_2$Te$_4$.  (a) Overview of septuple layer (SL) crystal structure with spins on Mn atoms indicated by red arrows. (b) DFT slab calculation of surface-projected band structure shown as black curves with the weight of the projected bands indicated by purple circles.  (c)-(e) Photoemission intensity (c) before the arrival of the pump pulse ($\Delta t < 0$) and (d) in the initial stage of excitation ($\Delta t = 40$~fs) with a laser fluence of 280~$\mu$J/cm$^2$.  (e) Intensity difference spectrum obtained by subtracting the equilibrium spectrum in (c) from the excited state signal in (d).  The spectra were collected around the $\Gamma$-point of the Brillouin zone.  The green and black arrows in (b) and (c) indicate the surface and bulk bands,  respectively.  The dashed green lines in (c) indicate the expected dispersion of the topological surface state. (f) Absolute value of normalised intensity difference integrated within the $(E,k)$-regions demarcated by correspondingly coloured boxes in the inset as a function of time delay. (g) Normalised intensity difference integrated within $(E,k)$-regions demarcated by the coloured boxes centred at varying energies placed along the most intense excited band, as shown in the inset. The intensity difference in the inset in (f)-(g) is the same as shown in (e). }
	\label{fig1}
\end{figure}

Figure \ref{fig1}(a) presents the SL MnBi$_2$Te$_4$ structure which can be thought of as a QL with an intercalated Mn-Te bilayer, introducing a plane of unpaired spins.  Non-trivial topology in this system derives from its Te~5$p$ valence band (VB) and Bi~6$p$ conduction band (CB) which undergo band inversion due to spin-orbit coupling (SOC) \cite{Li2019}.  The corresponding surface-projected electronic structure obtained from density functional theory (DFT) slab calculations is presented in Fig.~\ref{fig1}(b). The surface and bulk states are indicated by green and black arrows, respectively. In the energy range of our measurement, three branches of bulk bands and two surface bands are observed with the order from lower to higher energy of surface-bulk-bulk-surface-bulk around the $\Gamma$-point of the Brillouin zone. The two branches of bulk bands with lower energy exhibit weak out-of-plane-dispersion \cite{Estyunin2020,Yan2021}. 

Figures \ref{fig1}(c)-(d) present the electronic dispersion of MnBi$_2$Te$_4$ around the $\Gamma$-point of the Brillouin zone, measured in equilibrium conditions before optical excitation ($\Delta t < 0$) and during the initial moments of excitation ($\Delta t = 40$~fs).  Figure~\ref{fig1}(e) shows the photoemission intensity difference, resulting from subtraction of the equilibrium spectrum from the excited spectrum.  The red (blue) regions correspond to gain (loss) of photoemission signal, and can be interpreted as excited electron (hole) populations.  The spectra display a broad, nearly parabolic, distribution of intensity with a minimum at $E-E_F = -0.20$~eV,  indicating the as-grown crystals are significantly electron-doped \cite{Zeugner2019}.  Note the asymmetric presentation of the $k_{\parallel}$-scale is due to the fact that we have centred the excited state signal on the detector window, and not the $\Gamma$-point, in order to fully capture the dynamics along the band above the Fermi energy $E_F$. A region of high intensity is concentrated at $\Gamma$ around $E-E_F = -0.05$~eV, which is consistent with the location of surface states marked by a green arrow in Fig. \ref{fig1}(b). The fainter background intensity, marked by a black arrow in Fig. \ref{fig1}(c), is primarily composed of bulk conduction band (BCB) states, as seen via the comparison with Fig. \ref{fig1}(b). During excitation we observe that the bands crossing $E_F$ become occupied up to $E-E_F = 0.20$~eV with high excited state signal above the central surface states. These features have previously been attributed to a Rashba-like spin splitting \cite{Estyunin2020},  suggesting a significant non-equilibrium occupation of spin-polarized carriers in this system. The topological surface state in the bulk band gap (below $E-E_F = -0.20$~eV), reported to be a gapless Dirac cone in previous works \cite{PhysRevX.9.041038,PhysRevX.9.041039,PhysRevX.9.041040}, is not resolved due to the low cross-section compared with the CB states at our probe polarization and photon energy \cite{Yan2021}.

The time-dependence of electron (hole) dynamics in the BCB and surface states is examined by integrating the intensity within an $(E,k)$-region spanning 40~meV and 0.03~Å$^{-1}$, respectively,  centred $\pm$40~meV above (below) $E_F$.  These integration regions are placed at different $k$-values for electrons and holes in order to track the intensity difference for regions composed mainly of BCB (intensity away from $\Gamma$) or surface bands (intensity around $\Gamma$). Figure~\ref{fig1}(f) presents the resulting time-dependent intensity around $E_F$ for electrons, represented by red and orange curves, as well as holes, represented by light and dark blue curves that are color-coded according to the integration regions in the inset.  The intensity due to excited carriers initially matches the pump-probe cross-correlation integral (see dashed peak in Fig.~\ref{fig1}(f)).  The increase in signal then slows down before reaching a plateau, which is followed by a decay back to equilibrium that is achieved around 5~ps after the excitation.  The intensity difference corresponding to excited holes at -40~meV reaches its maximum at $\Delta t \approx 0.2$~ps. For excited electrons at 40~meV, the maximum excitation density is achieved significantly later, at $\Delta t \approx 0.8$~ps.  The $E$-dependent decay dynamics is determined by inspecting the intensity difference integrated in $(E,k)$-regions tracking the intense surface band, as shown in Fig.~\ref{fig1}(g). The signal furthest from $E_F$ rapidly increases during excitation and then decays exponentially while the transient population of excited electrons closer to $E_F$ builds up more slowly. This behavior is consistent with a cascading of intraband scattering processes from higher-lying states \cite{Hajlaoui2012, Sobota2013}.  The faster transient build-up of holes close to $E_F$ observed in Figs.~\ref{fig1}(f)-(g) reveals an apparent asymmetry in electron and hole dynamics, which is indicative of a shift of the chemical potential induced by the photodoping \cite{Crepaldi2012}.  Such a shift is demonstrated in Supplementary Figure S1. The slower decay dynamics of excited electrons and holes close to $E_F$ is driven by the evolution of the electronic temperature and associated Fermi-Dirac distribution after excitation when a quasi-thermal equilibrium is attained  \cite{Crepaldi2012}.

\begin{figure} [t!]
	\includegraphics[width=0.98\textwidth]{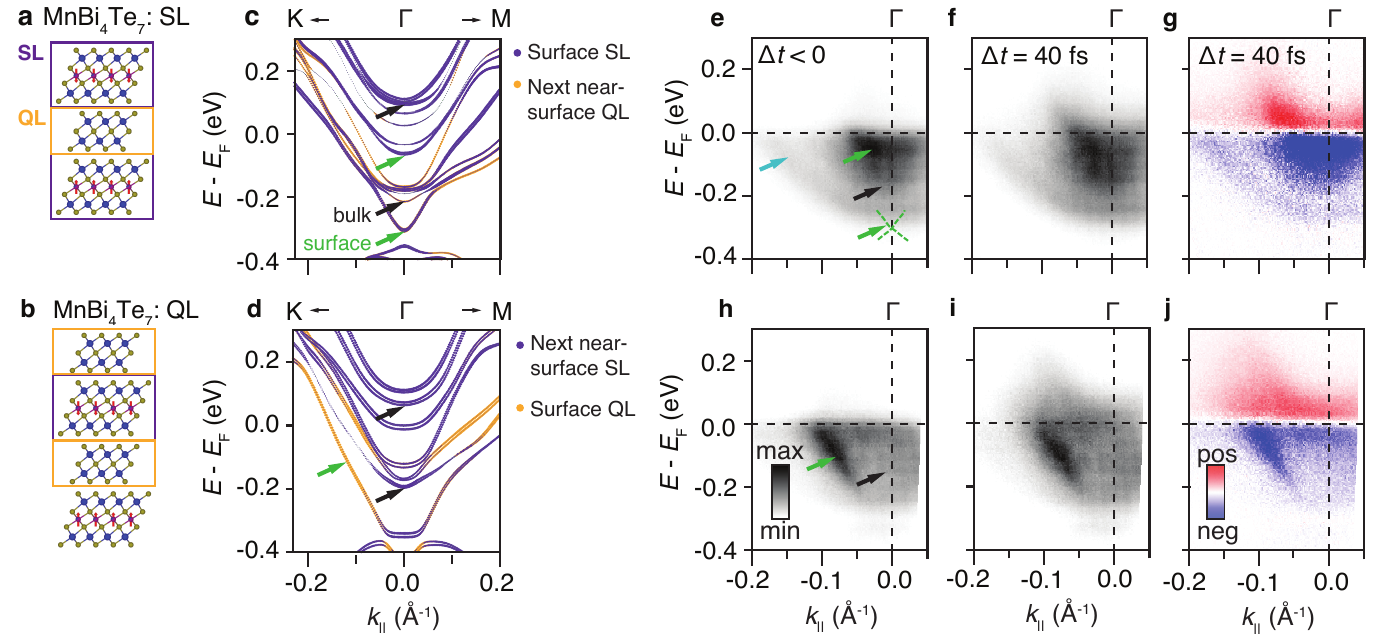}\\
\caption{Surface termination dependent electron dynamics in MnBi$_4$Te$_7$.  (a)-(b) Layered crystal structures of MnBi$_4$Te$_7$ with a SL or QL layer exposed at the surface. The red arrows indicate spins on the Mn atoms.  (c)-(d) Dispersions obtained from DFT slab calculations for MnBi$_4$Te$_7$ with (c) SL and (d) QL termination. The SL projected bands are colored purple while the QL projected bands are colored orange. (e)-(j) TR-ARPES spectra and intensity difference for (e)-(g) SL termination and (h)-(j) QL termination. The data are presented similarly as for MnBi$_2$Te$_4$ in Figs. \ref{fig1}(c)-(e). The measurements were performed with a laser fluence of 280~$\mu$J/cm$^2$. Arrows in (c)-(e) and (h) indicate surface (green), bulk (black) and surface-induced Rashba (light blue) bands.  Dashed green lines in (e) indicate the expected dispersion of the topological surface state.}
	\label{fig2}
\end{figure}

We now turn to the MnBi$_4$Te$_7$ heterostructure composed of alternating SL and QL units as sketched in Figs.~\ref{fig2}(a)-(b). Cleaving the bulk crystal can expose either of these terminations, which exhibit distinct surface electronic structures that we can separately resolve in our TR-ARPES experiment, as shown in Supplementary Figure S2.  Bulk- and surface-derived bands are distinguished via DFT slab calculations of the surface-projected band structures associated with the two terminations, as shown in Figs.~\ref{fig2}(c)-(d).  The calculated dispersions exhibit a gapped Dirac cone for the SL surface and hybridization bands between the two layers near the QL surface in agreement with previous studies \cite{wu2020distinct}. TR-ARPES spectra acquired within the same energy-range for both terminations are presented in Figs~\ref{fig2}(e)-(j). Both terminations exhibit a similar level of electron-doping as MnBi$_2$Te$_4$, causing the photoemission intensity to mainly consist of contributions from BCB and surface states around $E_F$.  Regions of high intensity measured at the SL termination do not coincide with regions of high intensity at the QL termination due to the surface-specific character of these bands (see green arrows in Figs.~\ref{fig2}(e) and \ref{fig2}(h)).  This assignment of band character is supported by the calculations for the SL termination that show bulk bands concentrated in the range of 0.1-0.2 eV below $E_F$ around $\Gamma$ whereas the bands immediately below $E_F$ are surface states.  Regions of high intensity for the QL termination originate from hybridized surface bands.  For both SL and QL terminations, the fainter background signals are consistent with BCB states that are identified via black arrows in Figs.  \ref{fig2}(c)-(d) and Figs.~\ref{fig2}(e) and \ref{fig2}(h).  Finally, the SL termination supports a surface-induced Rashba band marked by a light-blue arrow in Fig.  \ref{fig2}(e), which is not reproduced by DFT calculations \cite{Wu2020, Vidal2021}.  

The spectra acquired for $\Delta t = 40$~fs and the intensity difference calculated by subtracting the corresponding spectra measured for  $\Delta t < 0$ in Figs.  \ref{fig2}(f)-(g) and  \ref{fig2}(i)-(j) reveal significant electron and hole populations in the surface band around $\Gamma$ at the SL termination, whereas the QL termination exhibits a strong signal due to holes in the surface band, and mainly excited electrons in the bulk states above $E_F$ upon photoexcitation.The time-dependence of the intensity difference in selected $(E,k)$-regions is analyzed for the two terminations in Supplementary Figure~S3, similarly as shown for MnBi$_2$Te$_4$ in Figs. \ref{fig1}(f)-(g). The asymmetry between excited electron and hole signals driven by chemical potential shift diminishes from MnBi$_2$Te$_4$ to SL-terminated MnBi$_4$Te$_7$ and is absent in QL-terminated MnBi$_4$Te$_7$, consistent with significant interlayer charge transfer in MnBi$_4$Te$_7$.

\begin{figure} [t!]
	\includegraphics[width=1\textwidth]{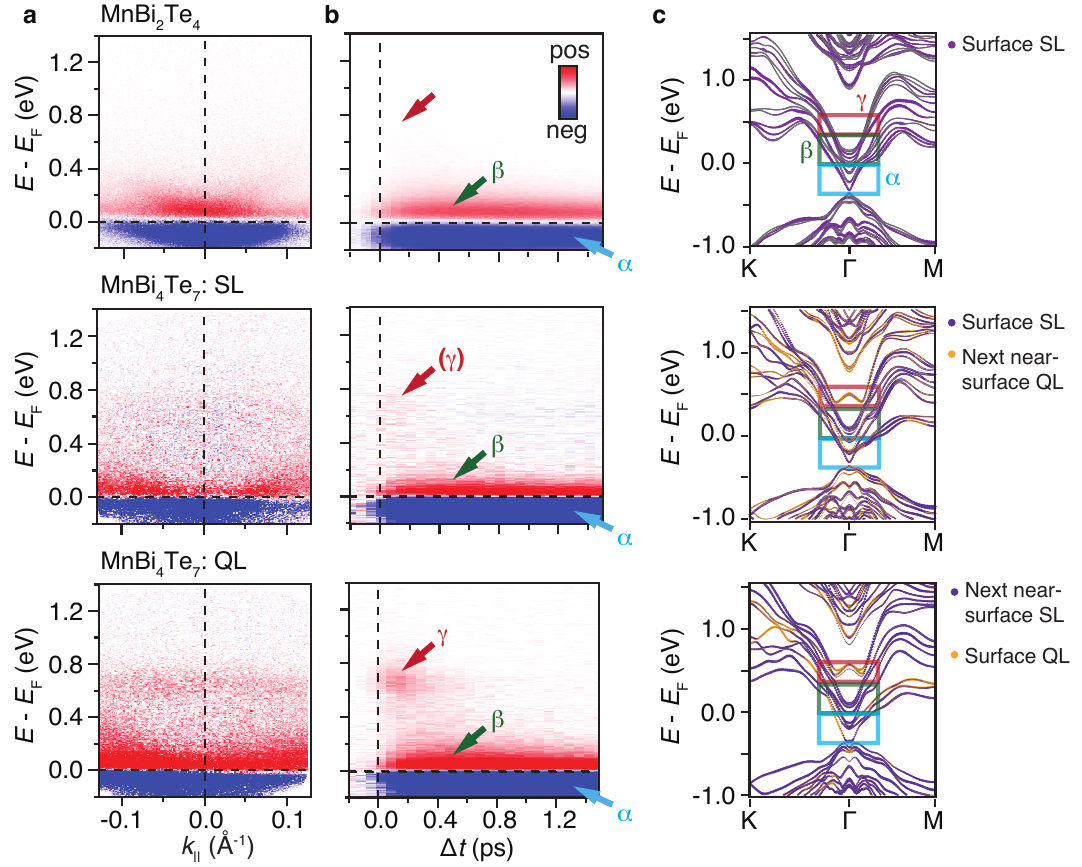}\\
\caption{Intra- and interband dynamics in the van der Waals heterostructures MnBi$_2$Te$_4$ and MnBi$_4$Te$_7$.  (a) Intensity difference spectra of MnBi$_2$Te$_4$ and SL and QL terminations of MnBi$_4$Te$_7$ obtained by subtracting an average of the equilibrium spectrum measured at $\Delta t < 0$ from the excited state signal at $\Delta t = 40$ fs. (b) Energy-distribution curves (EDCs) extracted at $\Gamma$ as a function of time delay for MnBi$_2$Te$_4$ (top), SL (middle) and  QL (bottom) terminations of MnBi$_4$Te$_7$.  The average laser fluence is 300 $\mu$J/cm$^2$ for the datasets presented here. (c) DFT-calculated slab bands for the structures indicated in (a) in the same row. The colored boxed areas in (c) and correspondingly colored arrows in (b) indicate regions of distinct dynamics labelled as $\alpha$, $\beta$ and $\gamma$.}
	\label{fig3}
\end{figure}

Having established the dispersion and response to photoexcitation across MnBi$_2$Te$_4$ and both MnBi$_4$Te$_7$ surface terminations,  we are now in a position to disentangle the layer-dependent contributions to the dynamics of the van der Waals-stacked MnBi$_2$Te$_4$ and Bi$_2$Te$_3$ layers of MnBi$_4$Te$_7$.  In Fig.~\ref{fig3}(a) we inspect the transient occupation of both terminations via the intensity difference at $\Delta t = 40$~fs and compare with MnBi$_2$Te$_4$ in a wider energy window in order to determine where the initially excited electrons are generated. The QL spectrum is characterized by a strong excitation signal 0.7~eV above $E_F$ while only extremely faint intensity is present in this region at the SL termination, likely emerging from the next-nearest QL layer. These signals are consistent with spatially-resolved pump-probe ARPES spectra obtained on the two terminations using a micro-focused beam \cite{Yan:2021_sciinst}. In analogy with Bi$_2$Te$_3$, the shallow excited band may be a surface resonance in the projected bulk band gap \cite{SanchezBarriga2017}.  Indeed, our slab band structure calculations presented over a wider energy range in Fig. \ref{fig3}(c) indicate these are QL-derived bands that are absent in the MnBi$_2$Te$_4$ system, as seen via the region enclosed by a red box in Fig. \ref{fig3}(c). 

Energy distribution curves (EDCs) extracted at $\Gamma$ as a function of time-delay, displayed in Fig.~\ref{fig3}(b), reveal that the band at 0.7~eV in the QL is instantaneously occupied followed by a rapid depopulation of excited electrons.  Concomitantly, the electron and hole signals grow in the states closer to $E_F$ in both the QL and SL. The infrared pump pulse initially generates electron-hole pairs via direct transitions from the initially occupied states in both MnBi$_2$Te$_4$ and MnBi$_4$Te$_7$ \cite{Hajlaoui2012, Sobota2013}.  These include the states labelled $\alpha$ in Fig.~\ref{fig3},  as well as states at lower energy in the VB manifold. In MnBi$_2$Te$_4$, where the QL-derived state at 0.7 eV is absent, excitation proceeds via high-lying bulk states followed by scattering directly to states close to $E_F$.  See Supplementary Figures S4-S5 for further details on the bulk band structures and symmetries of the high-lying states. In MnBi$_4$Te$_7$,  we propose that a fraction of excited carriers rapidly occupy the state at 0.7~eV (labelled $\gamma$) in the QLs of the van der Waals stack (on a faster time-scale than our experiment can resolve). Carriers are then extracted to the unoccupied states (labelled $\beta$) above $E_F$ in both the QL and SL via interband charge transfer with a timescale given by $\tau_0$.  A cascade of intraband scattering then results in excited carriers trickling down these states and finally recombining with holes via electron-phonon coupling on a timescale labeled as $\tau_1$.

\begin{figure} [t!]
	\includegraphics[width=1\textwidth]{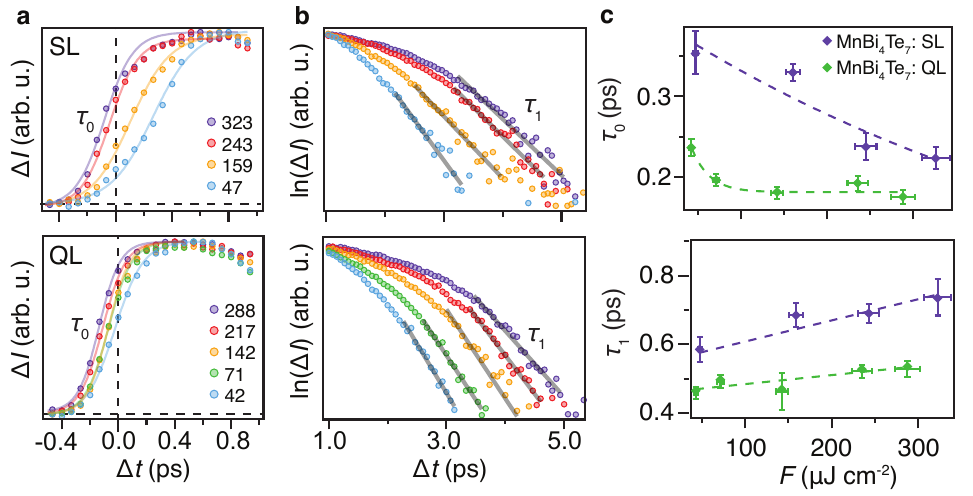}\\
\caption{Charge transfer and phonon relaxation in MnBi$_4$Te$_7$.  (a)-(b) $(E,k)$-integrated intensity difference from 0 to 70~meV above $E_F$ and from -0.2 to 0.2~Å$^{-1}$ for SL and QL terminations of MnBi$_4$Te$_7$.  Each data set was acquired with the fluence stated in (a) in units of $\mu$J/cm$^2$. In (a) we examine the rising edge of the signal (circles),  representing the filling of unoccupied states above $E_F$, and fit a sigmoid function (curves) to extract the timescale $\tau_0$ of the filling.  In (b) we analyze the decay part of the signal on a logarithmic intensity scale and perform exponential fits (lines) to extract the phonon relaxation timescale $\tau_1$.  (c) Extracted timescales from (a)-(b) as a function of laser fluence. Dashed curves provide a guide to the eye. }
	\label{fig4}
\end{figure}

The optical tunability of inter- and intralayer charge transfer and phonon relaxation between QL and SL layers of MnBi$_4$Te$_7$ is demonstrated in Fig~\ref{fig4}. We analyze the $(E,k)$-integrated intensity difference in the initially unoccupied states within 70~meV of $E_F$ as a function of laser fluence and consider separately the rise part of the signal in Fig.~\ref{fig4}(a) and the decay part on a logarithmic intensity scale in Fig.~\ref{fig4}(b). The initial increase of the signal in Fig.~\ref{fig4}(a) is highly fluence-dependent in the SL, exhibiting a significantly slower timescale with decreasing fluence, whereas the QL rise signal is faster and less sensitive to the fluence. In all cases, the rising part of the signal is well-described by a fit to a sigmoid function given by $(1 + \exp(2.35 \Delta t /\tau_0))^{-1}$, providing an estimate of the charge transfer time, $\tau_0$. After the initial increase in intensity, the excited state population enters a plateau region. This occurs due to the competition between the charge being transferred from the $\gamma$-state in the QL into the $\beta$-states in both the SL and QL and their subsequent recombination with holes. Finally, when the rate of refilling becomes slower than the electron-hole recombination processes, the transient occupation diminishes exponentially as shown in Fig.~\ref{fig4}(b). We quantify the timescale $\tau_1$ of this process via the slope of a linear fit to the logarithm of the trace in Fig.~\ref{fig4}(b).

The extracted fluence dependence of $\tau_0$ and $\tau_1$ is shown in Fig.~\ref{fig4}(c). The growth rate of the population in the SL reaches a maximum of $\tau_0 = 0.35$~ps at a low fluence of 47~$\mu$J/cm$^2$,  such that the population reaches its peak around $\Delta t = 0.80$~ps. This is substantially delayed compared to the peak being reached around $\Delta t = 0.40$~ps for $\tau_0 = 0.22$~ps at a high fluence of 323~$\mu$J/cm$^2$.  As we are considering the population in an energy range immediately above $E_F$, the rise time is strongly affected by interlayer charge transfer from the excited carriers in the $\gamma$-state in the QL. For low fluence, the intralayer filling of states from $\gamma$- to $\beta$-bands is more favourable than the interlayer charge transfer to the SL CB. Therefore the filling rate of $\beta$-bands in the SL is slower for low excitation densities. For high excitation density, on the other hand,  a large initial population is generated in the QL,  thereby removing this bottleneck effect.  For the QL termination, $\tau_0$ exhibits a less dramatic fluence-dependence and merely reflects the efficiency of intralayer interband scattering.

Finally, the decay time $\tau_1$ provides information on electron-hole recombination. By comparison to the related systems Bi$_2$Se$_3$ \cite{Sobota2013} and Bi$_2$Se$_x$Te$_{1-x}$ we infer that the relaxation is mediated by energy transfer to phonons. In particular,  terahertz spectroscopy studies \cite{Li2018, Reinhoffer2020} show similar fluence-dependent behavior of $\tau_1$. The decay constant increases with applied fluence, indicating that the electron-phonon coupling becomes less efficient with a larger population of excited carriers. This trend can be explained by increased screening of the electron-phonon interaction by the photoinduced charge carriers \cite{Heid2017}.
The relaxation dynamics extracted in Fig.~\ref{fig4}(c) are notably faster for QL than SL, suggesting that intralayer phonons play a dominant role in the scattering processes. Such behavior is in line with a substantially weakened c-axis interlayer vibrational interaction in the heterostructure \cite{Cho2022} in comparison to its non-magnetic counterpart Bi$_2$Te$_3$ \cite{Liang2017}. We note, however, that we cannot distinguish between the involvement of in-plane and out-of-plane intralayer vibrational modes in the scattering process.

In conclusion, we have determined the ultrafast charge carrier dynamics of intrinsic magnetic topological insulator van der Waals heterostructures MnBi$_2$Te$_4$ and MnBi$_4$Te$_7$.  The out-of-equilibrium distribution of excited carriers exhibit distinct dynamics that is strongly affected by interlayer interactions between the Bi$_2$Te$_3$ quintuple layers and the MnBi$_2$Te$_4$ septuple layers.  Infrared optical pumping leads to a transient population of carriers in the Bi$_2$Te$_3$ layers.  These excited carriers are then extracted via interband and interlayer scattering into conduction band states around the Fermi energy within the Bi$_2$Te$_3$ and in the adjacent MnBi$_2$Te$_4$ layers, respectively.  In the regime of low laser fluence,  the charge transfer processes into the MnBi$_2$Te$_4$ layers exhibit a bottleneck,  potentially enabling optical control of the population of spin-polarized carriers in the magnetic layers of the van der Waals stack.  Furthermore,  as the subsequent electron-phonon mediated relaxation is less efficient in the magnetic MnBi$_2$Te$_4$ compared to the non-magnetic Bi$_2$Te$_3$ layers it might be feasible to optically modify the magnetic  properties of the heterostructure when designing spintronic and optoelectronic devices with layers of MnBi$_2$Te$_4$.

\section{Methods}
\subsection{Sample growth}
The MnBi$_2$Te$_4$ and MnBi$_4$Te$_7$ crystals were grown by the flux method. High purity Mn, Bi, and Te powders were sealed in quartz tubes with a ratio of Mn:Bi:Te~=~1:10:16. The tubes were heated to 900$\degree$C and then cooled to 595/590$\degree$C, respectively. The flux was removed by centrifuging after cooling.

\subsection{Photoemission experiments}
The TR-ARPES measurements were performed at the T-ReX facility (Trieste, Italy). Pump and probe pulses were generated by a 250 kHz Ti:sapphire Coherent Reg A laser system, whose output was centered at an energy of 1.55~eV. The fourth harmonic (6.2~eV) of the fundamental beam was used as an $s$-polarised probe pulse. The remainder of the beam was used as a $p$-polarised pump with a tunable fluence in the range of 40-280~$\mu$J/cm$^2$, arriving at the sample at a variable time delay, $\Delta t$, with respect to the probe. The experimental time, energy and angular resolution were better than 200~fs, 50~meV and 0.2$\degree$, respectively. The samples were cleaved \textit{in situ} at a base pressure of 2$\cdot 10^{-10}$ mbar at room temperature, and kept at a temperature of 100-110~K during the measurement.

\subsection{Density functional theory calculations}
Band structures of MnBi$_2$Te$_4$ and MnBi$_4$Te$_7$ were calculated by DFT \cite{Hohenberg1964, Kohn1965}, which was carried out by the Vienna ab initio simulation package (VASP) \cite{Kresse1996} based on the projector augmented wave (PAW) method \cite{Kresse1999}. The exchange-correlation functional was described by the generalized gradient approximation with the Perdew-Burke-Ernzerhof formalism (PBE) \cite{Perdew1996} with on-site Coulomb interaction Hubbard $U = 5$~eV for electrons on d-orbitals of Mn atoms. The total energy convergence criteria was set to 1.0$\cdot 10^{-6}$~eV including spin-orbit coupling (SOC). The plane-wave cutoff energy was set to 400~eV. We employed an $11\times11\times1$ Monkhorst-Pack grid to sample the whole Brillouin-zone for both of the slab structures.  Our slab band structures of MnBi$_2$Te$_4$ were obtained from slab calculations with the thickness of 6 van der Waals layers, while that of QL and SL terminations of MnBi$_4$Te$_7$ employed 5 and 7 van der Waals layers, respectively. The magnetic configurations were A-type AFM for both cases with 4.6~$\mu$B magnetic moments on Mn atoms.

\section{Acknowledgement}
We gratefully acknowledge funding from VILLUM FONDEN through the Young Investigator Program (Grant. No. 15375), Villum Investigator Program (Grant. No. 25931) and the Centre of Excellence for Dirac Materials (Grant. No. 11744), and the Danish Council for Independent Research, Natural Sciences under the Sapere Aude program (Grant No. DFF-9064-00057B).  Work at Advanced Institute for Materials Research has benefited from support of WPI-AIMR, JSPS KAKENHI Basic Science A (18H03858), New Science (18H04473 and 20H04623), and Tohoku University FRiD program.  C.S. acknowledges Marie Sklodowska-Curie Postdoctoral Fellowship (proposal number 101059528). Work at SUSTech was supported by National Key R\&D Program of China under Grant No. 2020YFA0308900,and Center for Computational Science and Engineering of Southern University of Science and Technology.

\begin{suppinfo}
See Supporting Information at for 5 Sections containing 5 Figures describing the chemical potential shift induced by the pump pulse, the method to resolve the two terminations of MnBi$_4$Te$_7$ in the TR-ARPES experiment,  the $E$- and $k$-dependent dynamics of MnBi$_4$Te$_7$,  bulk band structure calculations of MnBi$_2$Te$_4$ and MnBi$_4$Te$_7$, and the symmetry of unoccupied states.  This material is available free of charge via the internet at \url{https://pubs.acs.org/doi/10.1021/acs.nanolett.2c03075} .
\end{suppinfo}

\providecommand{\latin}[1]{#1}
\makeatletter
\providecommand{\doi}
  {\begingroup\let\do\@makeother\dospecials
  \catcode`\{=1 \catcode`\}=2 \doi@aux}
\providecommand{\doi@aux}[1]{\endgroup\texttt{#1}}
\makeatother
\providecommand*\mcitethebibliography{\thebibliography}
\csname @ifundefined\endcsname{endmcitethebibliography}
  {\let\endmcitethebibliography\endthebibliography}{}

\end{document}